# Information Flow in Entangled Quantum Systems

## David Deutsch and Patrick Hayden

Centre for Quantum Computation
The Clarendon Laboratory
University of Oxford, Oxford OX1 3PU, UK

*All information in quantum systems is, notwithstanding Bell's theorem, localised. Measuring or otherwise interacting with a quantum system $\mathfrak{S}$ has no effect on distant systems from which $\mathfrak{S}$ is dynamically isolated, even if they are entangled with $\mathfrak{S}$. Using the Heisenberg picture to analyse quantum information processing makes this locality explicit, and reveals that under some circumstances (in particular, in Einstein-Podolski-Rosen experiments and in quantum teleportation) quantum information is transmitted through 'classical' (i.e. decoherent) information channels.*

1. **Quantum information**

It is widely believed (see e.g. Bennett and Shor (1998)) that in general a complete description of a composite quantum system is not deducible from complete descriptions of its subsystems unless the 'description' of each subsystem $\mathfrak{S}$ depends on what is going on in other subsystems from which $\mathfrak{S}$ is dynamically isolated. If this were so, then in quantum systems information would be a nonlocal quantity – that is to say, the information in a composite system would not be deducible from the information located in all its subsystems and, in particular, changes in the distribution of information in a spatially extended quantum system could not be understood wholly in terms of *information flow, i.e.* in terms of subsystems carrying information from one location to another. In this paper we shall show that this belief is false. It has given rise to a wide range of misconceptions, some of which we shall also address here, but our main concern will be with the analysis of information flow in quantum information-processing systems.

Any quantum 'two-state' system such as the spin of an electron or the polarisation of a photon can in principle be used as the physical realisation of a *qubit* (quantum bit), the basic unit of quantum information. When used to store or transmit discrete data, such as the values of integers, to an unknown destination, the capacity of a qubit is exactly one bit – in other words, it can hold one of two possible values; moreover, any observer who knows which of the qubit's observables the value was stored in can discover the value by measuring that observable. However, the states in which the qubit 'holds a value' in that sense are merely an isolated pair in a continuum of possible states. Hence there is a lot more than one bit of information in a qubit, though most of it is not accessible through measurements on that qubit alone. For a variety of theoretical and practical reasons, the study of the properties of this quantum information has recently been the subject of increasing attention (for a review, see Bennett and Shor (*loc. cit.*)). The main question we are addressing here is whether it possible to characterise such information *locally*, *i.e.* in such a way that a complete description of a composite system *can* always be deduced from complete descriptions of its subsystems, where under those descriptions, 'the real factual situation of the system $S_2$ is independent of what is done with the system $S_1$, which is spatially separated from the former' (Einstein (1949, p85)).

Einstein originally proposed this criterion during his celebrated debate with Bohr on the foundations of quantum theory, in which they both agreed that it is not satisfied by quantum theory. Bohr drew the lesson that there can be no such thing as 'the real factual situation of the system' except at the instant of measurement. Einstein concluded instead that quantum theory is incomplete and needs to be completed, perhaps by what we should now call a hidden-variable theory. Subsequent developments such as Bell's theorem (Bell (1964)) and Aspect's experiment (Aspect *et al.* (1982)), which are *prima facie* refutations of Einstein's conclusion, have therefore been taken as vindications of Bohr's. In fact, both conclusions are mistaken, having been drawn from the same false premise: as we shall show in this paper, quantum physics is entirely consistent with Einstein's criterion.

Our method is to consider a quantum system prepared in a way that depends on one or more parameters, and then to investigate where those parameters subsequently appear in descriptions of that system and others with which it interacts. Although we shall express our results in terms of the location and flow of information, we shall not require a quantitative definition of information. We require only that a system $\mathfrak{S}$ be deemed to *contain information* about a parameter θ if (though not necessarily only if) the probability of some outcome of some measurement on $\mathfrak{S}$ alone depends on θ; and that $\mathfrak{S}$ be deemed to *contain no information* about θ if there exists a complete description of $\mathfrak{S}$ that satisfies Einstein's criterion and is independent of θ.

## 2. Quantum theory of computation in the Heisenberg picture

Consider a quantum computational network $\mathfrak{N}$ containing *n* interacting qubits $\mathfrak{Q}_1,\ldots,\mathfrak{Q}_n$. Following Gottesman (1998), we may represent each qubit $\mathfrak{Q}_a$ at time *t* in the Heisenberg picture by a triple

$$\hat{\mathbf{q}}_a(t) = \left(\hat{q}_{ax}(t), \hat{q}_{ay}(t), \hat{q}_{az}(t)\right) \qquad (1)$$

of $2^n \times 2^n$ Hermitian matrices representing observables of $\mathfrak{Q}_a$, satisfying

$$\left. \begin{array}{ll} \left[\hat{\mathbf{q}}_a(t), \hat{\mathbf{q}}_b(t)\right] = 0 & (a \neq b), \\ \hat{q}_{ax}(t)\,\hat{q}_{ay}(t) = i\hat{q}_{az}(t) & \text{(and cyclic permutations over } (x,y,z)). \\ \hat{q}_{ax}(t)^2 = \hat{1} & \end{array} \right\} \qquad (2)$$

Thus each $\hat{\mathbf{q}}_a(t)$ is a representation of the Pauli spin operators $\hat{\boldsymbol{\sigma}} = (\hat{\sigma}_x, \hat{\sigma}_y, \hat{\sigma}_z)$, but in terms of time-dependent $2^n \times 2^n$ matrices instead of the usual constant $2 \times 2$ ones:

$$\hat{\sigma}_x = \begin{pmatrix} 0 & 1 \\ 1 & 0 \end{pmatrix}, \quad \hat{\sigma}_y = \begin{pmatrix} 0 & -i \\ i & 0 \end{pmatrix}, \quad \hat{\sigma}_z = \begin{pmatrix} 1 & 0 \\ 0 & -1 \end{pmatrix}. \tag{3}$$

We may choose, as the computation basis at time $t$, the simultaneous eigenstates $\{|z_1,\ldots,z_k;t\rangle\}$ of the $\{\hat{z}_a(t)\}$, where

$$\hat{z}_a(t) = \tfrac{1}{2}\left(\hat{1} + \hat{q}_{az}(t)\right). \tag{4}$$

Each $\hat{z}_a(t)$ has eigenvalues 0 and 1 (corresponding respectively to the eigenvalues $-1$ and $+1$ of $\hat{q}_{az}(t)$) and is the projector for the $a$'th qubit to hold the value 1 at time $t$.

There is considerable freedom in the choice of matrix representations for the observables (1). It is always possible, and usually desirable, to choose the initial representation to be

$$\hat{\mathbf{q}}_a(0) = \hat{1}^{a-1} \otimes \hat{\boldsymbol{\sigma}} \otimes \hat{1}^{n-a}, \tag{5}$$

where '$\otimes$' denotes the tensor product (distributed, in (5), over the three components of $\hat{\boldsymbol{\sigma}}$), and $\hat{1}^k$ is the tensor product of $k$ copies of the $2 \times 2$ unit matrix. As we shall see, once the qubits begin to interact, the observables immediately lose the form (5) in the original basis. That is because, as in classical mechanics, the value of each observable of one system becomes a function of the values of observables of other systems at previous times – though now the 'values' are matrices. (However, at every instant, because the conditions (2) are preserved by all quantum interactions, there exists a basis in which the observables take the form (5).)

The Heisenberg state of the network is of course constant and, in the theory of computation, it is often desirable to make it a *standard* constant $|0,\ldots,0;0\rangle$, so that the resources required to prepare the 'initial' state will automatically be taken into account in the analysis of computations. When studying algorithms whose intended inputs are qubits in unknown initial states, it may be convenient to work with other Heisenberg states $|\Psi\rangle \neq |0,\ldots,0;0\rangle$ but note, nevertheless, that by choosing any unitary matrix $\mathsf{U}$ with the property $|\Psi\rangle = \mathsf{U}|0,\ldots,0;0\rangle$, and setting $\hat{\mathbf{q}}_a(0) = \mathsf{U}^{\dagger}\left(\hat{1}^{a-1} \otimes \hat{\boldsymbol{\sigma}} \otimes \hat{1}^{n-a}\right)\mathsf{U}$ instead of (5), it is always *possible* to choose the Heisenberg state to be $|0,\ldots,0;0\rangle$.

The formalism presented here can be generalised to accommodate mixed states (see Deutsch *et al.* (1999)). That complication is unnecessary for present purposes, but note that even in the mixed state formalism it remains possible to choose the Heisenberg state to be $|0,\ldots,0;0\rangle$.

In what follows, we shall make that choice. For the sake of brevity, let us define

$$\langle \hat{A} \rangle \equiv \langle 0,\ldots,0;0|\hat{A}|0,\ldots,0;0\rangle \qquad (6)$$

for each observable $\hat{A}$ of $\mathfrak{N}$. Note that all predictions about the behaviour of $\mathfrak{N}$ can be expressed entirely in terms of expectation values of the form (6).

Let us assume for simplicity that each gate of $\mathfrak{N}$ performs its operation in a fixed period, and let us measure time in units of that period. The effect of a *k*-qubit gate **G** acting between the times *t* and *t*+1 is

$$\hat{\mathbf{q}}_a(t+1) = \mathsf{U}_{\mathbf{G}}^{\dagger}\left(\hat{\mathbf{q}}_{1'}(t),\ldots,\hat{\mathbf{q}}_{k'}(t)\right)\hat{\mathbf{q}}_{a'}(t)\mathsf{U}_{\mathbf{G}}\left(\hat{\mathbf{q}}_{1'}(t),\ldots,\hat{\mathbf{q}}_{k'}(t)\right), \qquad (7)$$

where $1',\ldots,k'$ are the indices of the qubits that are acted upon by **G**, and $a'$ is any such index. Since each qubit is acted upon by exactly one gate during any one computational step (counting the 'unit wire' **I**, which has no effect on the computational state of a qubit, as a gate with $\mathsf{U}_\mathbf{I} = \hat{1}$), the dynamical evolution of any qubit of $\mathfrak{N}$ during one step is fully specified by an expression of the form (7), where **G** is the gate acting on that qubit during that step. The form of each $\mathsf{U}_\mathbf{G}$ *qua* function of its arguments is fixed and characteristic of the corresponding gate **G**, and its form *qua* unitary matrix varies accordingly.

It follows that the simultaneous eigenstates of the $\{\hat{z}_{a'}(t)\}$ evolve according to

$$|z_{1'},\ldots,z_{k'};t+1\rangle = \mathsf{U}_{\mathbf{G}}^{\dagger}\big(\hat{\mathbf{q}}_{1'}(t),\ldots,\hat{\mathbf{q}}_{k'}(t)\big)|z_{1'},\ldots,z_{k'};t\rangle. \tag{8}$$

The computation basis evolves similarly, with *k* replaced by the total number of qubits *n*, and with $\mathsf{U}_{\mathbf{G}}$ replaced by the product (in any order, since they must commute) of all the unitary matrices corresponding to gates acting at time *t*.

We are now in a position to verify that quantum systems have the locality properties stated in Section 1. If we always choose the state vector to be a standard constant, the term 'state vector' becomes a misnomer, for the vector $|0,\ldots,0;0\rangle$ contains no information about the state of $\mathfrak{N}$ or anything else. All the information is contained in the observables. Specifically, the matrix triplets $\{\hat{\mathbf{q}}_a(t)\}$, each of which constitutes a complete (indeed redundant) description of one qubit $\mathfrak{Q}_a$, jointly constitute a complete description of the composite system $\mathfrak{N}$ – as promised.

As for Einstein's criterion about the effect of one subsystem upon another, consider a particular qubit $\mathfrak{Q}_a$ and let **F** be a gate that acts only on one or more qubits *other than* $\mathfrak{Q}_a$ (so that $\mathfrak{Q}_a$ is dynamically isolated from those qubits) during the period between *t* and *t*+1. According to (7), the complete description of $\mathfrak{Q}_a$ during that period would be unchanged if **F** were replaced by any other gate. Hence it is a general feature of this formalism that when a gate acts on any set of qubits, the descriptions of all other qubits remain unaffected – even qubits that are entangled with those that the gate acts on. This is, again, as promised.

A quantum computational network is not a general quantum system: for instance, its interactions all take place in discrete computational steps of fixed duration, and during any computational step each of its qubits interacts only with the other qubits that are acted upon by the same gate. Nevertheless, since every quantum system can be simulated with arbitrary accuracy by quantum computational networks (Deutsch 1989), the above conclusions about locality are true for general quantum systems too.

### 3. Some specific quantum gates

We often define gates according to the effect they are to have on the computation basis. In such cases we can use (8) to determine the form of the function $\mathsf{U}_\mathbf{G}$ associated with a given gate **G**. For example, a **not**-gate acting on a network consisting of a single qubit at time $t$ must have the effect

$$\left.\begin{array}{l}|0;t\rangle = |1;t+1\rangle = \mathsf{U}_{\mathbf{not}}^\dagger(\hat{\mathbf{q}}(t))|1;t\rangle \\ |1;t\rangle = |0;t+1\rangle = \mathsf{U}_{\mathbf{not}}^\dagger(\hat{\mathbf{q}}(t))|0;t\rangle\end{array}\right\}. \tag{9}$$

(Recall that the kets here are not Schrödinger states but eigenstates of Heisenberg observables. So, for instance, $|0;t\rangle$ in (9) is the zero-eigenvalue eigenstate of $\hat{z}(t) = \tfrac{1}{2}(\hat{1}+\hat{q}_z(t))$.) Hence at $t = 0$,

$$\langle r;0|\mathsf{U}_{\mathbf{not}}^\dagger(\hat{\mathbf{q}}(0))|s;0\rangle = \delta(r,1-s). \tag{10}$$

The Pauli matrices (3) together with the unit matrix form a basis in the vector space of all $2\times 2$ matrices, so we may express (10) as an expansion in this basis to obtain

$$\mathsf{U}_{\mathbf{not}}(\hat{\mathbf{q}}(0)) = \hat{\sigma}_x. \tag{11}$$

Using (5), (11) and the fact that the functional form of $\mathsf{U}_{\mathbf{not}}$ is constant, we infer that for a general network at a general time $t$, the unitary matrix associated with a **not**-gate acting on the $k'$th qubit is

$$\mathsf{U}_{\mathbf{not}k}(\hat{\mathbf{q}}_1(t),\ldots,\hat{\mathbf{q}}_n(t)) = \hat{q}_{kx}(t). \tag{12}$$

From (12) and (2) it follows that the effect of **not** on the $k'$th qubit is:

**'not'**: $\quad \hat{\mathbf{q}}_k(t+1) \equiv (\hat{q}_{kx}(t+1), \hat{q}_{ky}(t+1), \hat{q}_{kz}(t+1)) = (\hat{q}_{kx}(t), -\hat{q}_{ky}(t), -\hat{q}_{kz}(t))$, (13)

with all other qubits remaining unchanged, and from this we can immediately verify that the following operation on $\mathbb{Q}_k$:

**'√not'**: $\quad \hat{\mathbf{q}}_k(t+1) = (\hat{q}_{kx}(t), \hat{q}_{kz}(t), -\hat{q}_{ky}(t))$, (14)

is a 'square-root-of-not' operation (Deutsch (1987)).

Consider next the 'perfect-measurement' or *controlled-not* operation, **cnot** (Barenco *et al.* (1995)). This is an operation on two qubits, designated the *control* qubit and the *target* qubit. Its effect is that if the control qubit takes the value 0 then the target qubit is unaltered, and if the control qubit takes the value 1 then the target qubit is toggled. Given (12), this means that

$$\mathsf{U}_{\mathbf{cnot}}(\hat{\mathbf{q}}_k, \hat{\mathbf{q}}_l) = \hat{1}_k \otimes \frac{(\hat{1}_l - \hat{q}_{lz})}{2} + \hat{q}_{kx} \otimes \frac{(\hat{1}_l + \hat{q}_{lz})}{2}. \qquad (15)$$

where the *k*'th and *l*'th qubits are the 'target' and 'control' qubits respectively. Substituting (15) into (7), we obtain

**'cnot'**: $\quad \begin{Bmatrix} \hat{\mathbf{q}}_k(t+1) \\ \hat{\mathbf{q}}_l(t+1) \end{Bmatrix} = \begin{Bmatrix} (\hat{q}_{kx}(t), \ -\hat{q}_{ky}(t)\hat{q}_{lz}(t), \ -\hat{q}_{kz}(t)\hat{q}_{lz}(t)) \\ (\hat{q}_{kx}(t)\hat{q}_{lx}(t), \ \hat{q}_{kx}(t)\hat{q}_{ly}(t), \ \hat{q}_{lz}(t)) \end{Bmatrix}.$ (16)

Let $\mathbf{R_n}(\theta)$ be the single-qubit gate that would, if the *k*'th qubit were a spin-$\frac{1}{2}$ particle, rotate it through an angle $\theta$ about the unit 3-vector $\mathbf{n}$. The matrices $\hat{\mathbf{q}}_k$ must transform under this rotation in the same way as Pauli matrices do:

**'$R_n(\theta)$'**: $\quad \hat{\mathbf{q}}_k(t+1) = e^{i\frac{\theta}{2}\mathbf{n}\cdot\hat{\mathbf{q}}_k(t)} \hat{\mathbf{q}}_k(t) e^{-i\frac{\theta}{2}\mathbf{n}\cdot\hat{\mathbf{q}}_k(t)}.$ (17)

Hence in particular, the effect of rotating the *k*'th qubit through an angle θ about the *x*-axis is:

**'$R_x(\theta)$'**:     $\hat{\mathbf{q}}_k(t+1) = \left(\hat{q}_{kx}(t), \hat{q}_{ky}(t)\cos\theta + \hat{q}_{kz}(t)\sin\theta, \hat{q}_{kz}(t)\cos\theta - \hat{q}_{ky}(t)\sin\theta\right).$  (18)

Another useful gate, the 'Hadamard gate' **H**, is also a special case of (17), with $\theta = \pi$ and **n** bisecting the angle between the *x*- and *z*-axes:

**'H'**:     $\hat{\mathbf{q}}_k(t+1) = \left(\hat{q}_{kz}(t), -\hat{q}_{ky}(t), \hat{q}_{kx}(t)\right).$  (19)

In general, since the **cnot** gate together with gates of the type $\mathbf{R}_\mathbf{n}(\theta)$ constitute a universal set, the effect of any gate can be calculated by considering a computationally equivalent network containing only those gates, and then using (16) and (17).

For example, the gate that performs the so-called Bell transformation on two qubits (Braunstein *et al.* (1992)) is equivalent to the network shown on the right of the equals sign in Fig. 1. (Gates other than **cnot** are represented by rectangles, the vertical lines represent the paths of qubits, and the arrows at the top indicate their direction of motion.) Since both **cnot** and **H** are their own inverses, the same network upside-down (*i.e.* with **H** preceding **cnot**) performs

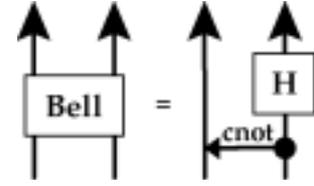

Fig. 1: The **Bell** Gate

the inverse of the Bell transformation. It follows that the effect of the **Bell** gate is

**'Bell'**:     $\begin{Bmatrix} \hat{\mathbf{q}}_k(t+1) \\ \hat{\mathbf{q}}_l(t+1) \end{Bmatrix} = \begin{Bmatrix} \left(\hat{q}_{kx}(t), -\hat{q}_{ky}(t)\hat{q}_{lz}(t), -\hat{q}_{kz}(t)\hat{q}_{lz}(t)\right) \\ \left(\hat{q}_{lz}(t), -\hat{q}_{kx}(t)\hat{q}_{ly}(t), \hat{q}_{kx}(t)\hat{q}_{lx}(t)\right) \end{Bmatrix},$  (20)

and the effect of its inverse is

**'Bell$^{-1}$'**:     $\begin{Bmatrix} \hat{\mathbf{q}}_k(t+1) \\ \hat{\mathbf{q}}_l(t+1) \end{Bmatrix} = \begin{Bmatrix} \left(\hat{q}_{kx}(t), -\hat{q}_{ky}(t)\hat{q}_{lx}(t), -\hat{q}_{kz}(t)\hat{q}_{lx}(t)\right) \\ \left(\hat{q}_{kx}(t)\hat{q}_{lz}(t), -\hat{q}_{kx}(t)\hat{q}_{ly}(t), \hat{q}_{lx}(t)\right) \end{Bmatrix}.$  (21)

## 4. Information flow in Einstein-Podolski-Rosen experiments

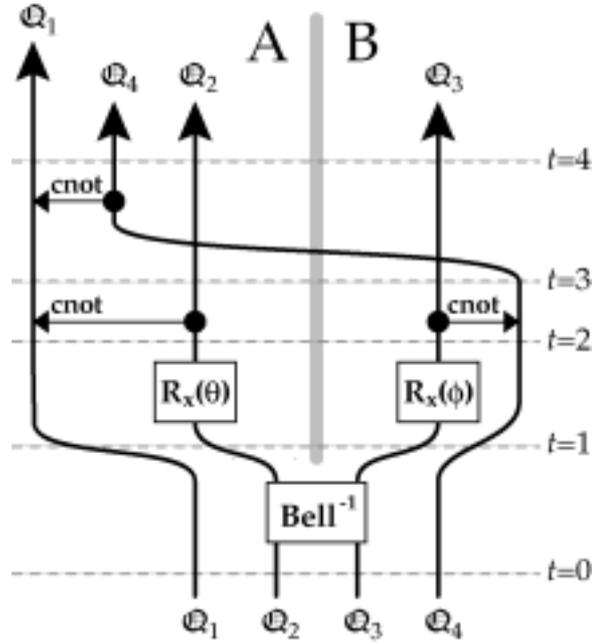

Fig. 2: An Einstein-Podolski-Rosen Experiment

The quantum computational network for performing an Einstein-Podolski-Rosen (EPR) experiment is shown in Fig. 2. Since this is the archetypal experiment that has been thought to demonstrate the nonlocal nature of information in quantum physics, it is instructive to trace the paths that information takes during the course of such an experiment. In particular, we shall trace how quantum information about the value of an angle $\phi$, chosen arbitrarily in a region B, reaches a distant region A.

Starting at time $t=0$ with four qubits $\mathbb{Q}_1\ldots\mathbb{Q}_4$ in the standard state $|0,0,0,0;0\rangle$, we entangle $\mathbb{Q}_2$ with $\mathbb{Q}_3$ by performing the inverse Bell operation (21). In Schrödinger-picture terminology they are now in the state

$$|\psi(1)\rangle = \tfrac{i}{\sqrt{2}}(|0\rangle|0\rangle - |1\rangle|1\rangle), \qquad (22)$$

but in the Heisenberg picture we have

$$\left.\begin{aligned}\hat{\mathbf{q}}_2(1) &= \hat{1}\otimes\left(\hat{\sigma}_x\otimes\hat{1},\ -\hat{\sigma}_y\otimes\hat{\sigma}_x,\ -\hat{\sigma}_z\otimes\hat{\sigma}_x\right)\otimes\hat{1} \\ \hat{\mathbf{q}}_3(1) &= \hat{1}\otimes\left(\hat{\sigma}_x\otimes\hat{\sigma}_z,\ -\hat{\sigma}_x\otimes\hat{\sigma}_y,\ \hat{1}\otimes\hat{\sigma}_x\right)\otimes\hat{1}\end{aligned}\right\}. \qquad (23)$$

After that (at $t=1$) we physically separate $\mathbb{Q}_1$ and $\mathbb{Q}_2$ from $\mathbb{Q}_3$ and $\mathbb{Q}_4$, moving these respective pairs to two regions A and B which are sufficiently far apart (or sufficiently isolated from each other) for nothing to be able to travel from either of them to the other until after $t=3$.

Then (still at $t=1$ as far as the computation is concerned, though in reality some time would be needed for the qubits to travel to A and B) we rotate $Q_2$ and $Q_3$ about their *x*-axes through arbitrarily (but locally) chosen angles $\theta$ and $\phi$ respectively. At this time ($t=2$), $Q_1$ and $Q_4$ have not yet participated in the computation and have therefore remained unchanged:

$$\hat{\mathbf{q}}_1(2) = \hat{\mathbf{q}}_1(0) = \hat{\boldsymbol{\sigma}} \otimes \hat{1}^3; \quad \hat{\mathbf{q}}_4(2) = \hat{\mathbf{q}}_4(0) = \hat{1}^3 \otimes \hat{\boldsymbol{\sigma}}, \tag{24}$$

but the descriptors of $Q_2$ and $Q_3$ are now functions of $\theta$ and $\phi$ respectively:

$$\left.\begin{aligned}\hat{\mathbf{q}}_2(2) &= \hat{1} \otimes \left(\hat{\sigma}_x \otimes \hat{1}, \ -\left(\cos\theta\hat{\sigma}_y + \sin\theta\hat{\sigma}_z\right) \otimes \hat{\sigma}_x, \ \left(\sin\theta\hat{\sigma}_y - \cos\theta\hat{\sigma}_z\right) \otimes \hat{\sigma}_x\right) \otimes \hat{1} \\ \hat{\mathbf{q}}_3(2) &= \hat{1} \otimes \left(\hat{\sigma}_x \otimes \hat{\sigma}_z, \ \sin\phi\hat{1} \otimes \hat{\sigma}_x - \cos\phi\hat{\sigma}_x \otimes \hat{\sigma}_y, \ \left(\cos\phi\hat{1} \otimes \hat{\sigma}_x + \sin\phi\hat{\sigma}_x \otimes \hat{\sigma}_y\right)\right) \otimes \hat{1}\end{aligned}\right\}. \tag{25}$$

Now, given the qualitative properties of information that we stated at the end of Section 1, and since, as we shall see, the values of $\theta$ and $\phi$ will affect the probabilities of the outcomes of measurements performed later in the experiment, we can infer that the system as a whole contains information about $\theta$ and $\phi$ at $t=2$. Furthermore, from (24) and (25) we know that none of the information about $\theta$ is contained in $Q_1$, $Q_3$ or $Q_4$, so we must conclude that it is located entirely in $Q_2$. Similarly, all the information about $\phi$ that is in the network at $t=2$ is located in $Q_3$. However, since all observables on $Q_2$ are linear combinations of the unit observable and the three components of $\hat{\mathbf{q}}_2(2)$, and since

$$\langle\hat{\mathbf{q}}_2(2)\rangle = (0,0,0) \tag{26}$$

is independent of $\theta$, the probability of any outcome of any possible measurement of any observable of $Q_2$ at $t=2$ is independent of $\theta$. In other words, the information about $\theta$, though present in $Q_2$, is not detectable by measurements on $Q_2$ alone.

Let us define *locally inaccessible information* as information which is present in a system but does not affect the probability of any outcome of any possible measurement on that system alone. We have shown that at $t=2$, the information about θ in $\mathbb{Q}_2$ is locally inaccessible, and so is the information about ϕ in $\mathbb{Q}_3$.

Nevertheless, such information can, and generically does, spread to other qubits through further local interactions. For example, in practice it spreads into the local environment through the unwanted interactions that cause decoherence. It also spreads to other qubits in our EPR experiment, where we now (after $t=2$) use **cnot** gates to perform perfect measurements on $\mathbb{Q}_2$ and $\mathbb{Q}_3$, recording the outcomes in $\mathbb{Q}_1$ and $\mathbb{Q}_4$ respectively. We then have:

$$\left. \begin{aligned} \hat{\mathbf{q}}_1(3) &= \left( \hat{\sigma}_x \otimes \hat{1}^2, \quad \hat{\sigma}_y \otimes \left(\cos\theta\hat{\sigma}_z - \sin\theta\hat{\sigma}_y\right) \otimes \hat{\sigma}_x, \quad \hat{\sigma}_z \otimes \left(\cos\theta\hat{\sigma}_z - \sin\theta\hat{\sigma}_y\right) \otimes \hat{\sigma}_x \right) \otimes \hat{1}, \\ \hat{\mathbf{q}}_2(3) &= \left( \hat{\sigma}_x \otimes \hat{\sigma}_x \otimes \hat{1}, \quad -\hat{\sigma}_x \otimes \left(\cos\theta\hat{\sigma}_y + \sin\theta\hat{\sigma}_z\right) \otimes \hat{\sigma}_x, \quad \hat{1} \otimes \left(\sin\theta\hat{\sigma}_y - \cos\theta\hat{\sigma}_z\right) \otimes \hat{\sigma}_x \right) \otimes \hat{1}, \\ \hat{\mathbf{q}}_3(3) &= \hat{1} \otimes \left( \hat{\sigma}_x \otimes \hat{\sigma}_z \otimes \hat{\sigma}_x, \quad \left(\sin\phi\hat{1} \otimes \hat{\sigma}_x - \cos\phi\hat{\sigma}_x \otimes \hat{\sigma}_y\right) \otimes \hat{\sigma}_x, \quad \left(\cos\phi\hat{1} \otimes \hat{\sigma}_x + \sin\phi\hat{\sigma}_x \otimes \hat{\sigma}_y\right) \otimes \hat{1} \right), \\ \hat{\mathbf{q}}_4(3) &= \hat{1} \otimes \left( \hat{1}^2 \otimes \hat{\sigma}_x, \quad -\left(\cos\phi\hat{1} \otimes \hat{\sigma}_x + \sin\phi\hat{\sigma}_x \otimes \hat{\sigma}_y\right) \otimes \hat{\sigma}_y, \quad -\left(\cos\phi\hat{1} \otimes \hat{\sigma}_x + \sin\phi\hat{\sigma}_x \otimes \hat{\sigma}_y\right) \otimes \hat{\sigma}_z \right) \end{aligned} \right\} \quad (27)$$

The locality of all these operations is reflected in the fact that at this time ($t=3$), $\hat{\mathbf{q}}_1$ and $\hat{\mathbf{q}}_2$ depend on θ but not ϕ, while $\hat{\mathbf{q}}_3$ and $\hat{\mathbf{q}}_4$ depend on ϕ but not θ. Again, it is easily verified that none of these dependences is detectable locally – *i.e.* by any measurement performed jointly on $\mathbb{Q}_1$ and $\mathbb{Q}_2$, or jointly on $\mathbb{Q}_3$ and $\mathbb{Q}_4$ – and that this would remain true if any amount of further interaction with other local qubits, or with the local environments, were to occur.

Finally, we measure whether the two outcomes that are now (at $t=3$) stored in $\mathbb{Q}_1$ and $\mathbb{Q}_4$ were the same or not. We do this by bringing $\mathbb{Q}_4$ (and thereby its information about ϕ) to location A and then using it as the control qubit of a **cnot** operation with $\mathbb{Q}_1$ as the target. The probability that the two outcomes were different is then $\langle \hat{z}_1(4) \rangle$. Using (4), (16), (27) and (2), we find that

$$\begin{aligned}\langle \hat{z}_1(4)\rangle &= \tfrac{1}{2}\langle \hat{1}+\hat{q}_{1z}(4)\rangle \\ &= \tfrac{1}{2}-\tfrac{1}{2}\langle \hat{q}_{1z}(3)\hat{q}_{4z}(3)\rangle \\ &= \tfrac{1}{2}-\tfrac{1}{2}\langle \hat{\sigma}_z \otimes \left(\cos\theta\cos\phi\,\hat{\sigma}_z\otimes\hat{1}-\cos\theta\sin\phi\,\hat{\sigma}_y\otimes\hat{\sigma}_z-\sin\theta\cos\phi\,\hat{\sigma}_y\otimes\hat{1}-\sin\theta\sin\phi\,\hat{\sigma}_z\otimes\hat{\sigma}_z\right)\otimes\hat{\sigma}_z\rangle \\ &= \cos^2\tfrac{1}{2}(\theta-\phi).\end{aligned} \qquad (28)$$

This is a familiar result, but in the course of calculating it in the Heisenberg picture, we have discovered exactly how the information about $\phi$ reached $\mathbb{Q}_1$: it was carried there in the qubit $\mathbb{Q}_4$ as it travelled from B to A.

It is easily verified that the result of the experiment would be unchanged if $\hat{z}_4(t)$ were measured any number of times on $\mathbb{Q}_4$'s journey from B to A. The locally inaccessible information about $\phi$ that is carried in $\mathbb{Q}_4$ would not be affected by such measurements, nor, therefore, would it be affected if $\mathbb{Q}_4$ suffered decoherence through environmental interactions that stabilised $\hat{z}_4$. But it would be copied into other qubits, and any qubit holding the outcome of a measurement of $\hat{z}_4$ could be used instead of $\mathbb{Q}_4$ to carry the information to A. The ability of quantum information to flow through a classical channel in this way, surviving decoherence, is also the basis of *quantum teleportation*, a remarkable phenomenon to which we now turn.

## 5. Information flow in quantum teleportation

The very term 'teleportation' was chosen by the discoverers of the phenomenon (Bennett *et al.* (1993)) because it was deemed to be a spectacular example of information from one location A appearing at another location B without being carried there in any physical object travelling from A to B – *i.e.* without information flow.

A quantum computational network for demonstrating teleportation is shown in Fig. 3. The information of interest is the angle θ through which, at $t = 0$, we choose to rotate the qubit $\mathbb{Q}_1$, located at A, about its *x*-axis. More generally, the $\mathbf{R}_x(\theta)$ gate could be replaced by an arbitrary single-qubit gate that prepared $\mathbb{Q}_1$ in an arbitrary pure state, which the network would 'teleport' to $\mathbb{Q}_5$ at the distant location B, but for simplicity we are restricting ourselves to a one-parameter family of states. We have

$$\hat{\mathbf{q}}_1(1) = \left( \hat{\sigma}_x, \quad \cos\theta\hat{\sigma}_y + \sin\theta\hat{\sigma}_z, \quad \cos\theta\hat{\sigma}_z - \sin\theta\hat{\sigma}_y \right) \otimes \hat{1}^4. \tag{29}$$

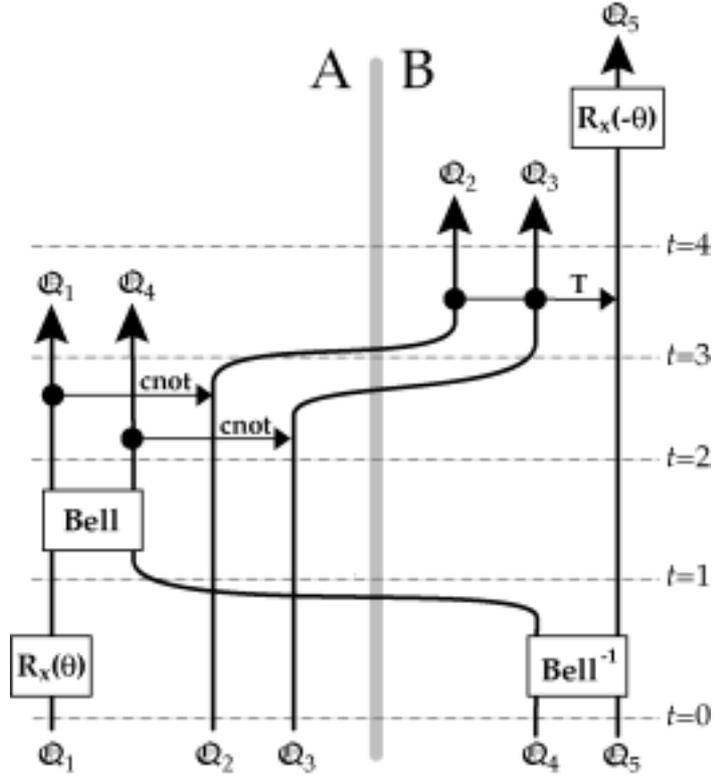

Fig. 3: Quantum Teleportation experiment

Also at $t = 0$, at location B, qubits $\mathbb{Q}_4$ and $\mathbb{Q}_5$ are entangled by the action of an inverse Bell gate:

$$\left. \begin{aligned} \hat{\mathbf{q}}_4(1) &= \hat{1}^3 \otimes \left( \hat{\sigma}_x \otimes \hat{1}, \quad -\hat{\sigma}_y \otimes \hat{\sigma}_x, \quad -\hat{\sigma}_z \otimes \hat{\sigma}_x \right) \\ \hat{\mathbf{q}}_5(1) &= \hat{1}^3 \otimes \left( \hat{\sigma}_x \otimes \hat{\sigma}_z, \quad -\hat{\sigma}_x \otimes \hat{\sigma}_y, \quad \hat{1} \otimes \hat{\sigma}_x \right) \end{aligned} \right\}. \tag{30}$$

Then the qubit $\mathbb{Q}_4$ travels to location A and undergoes a Bell operation together with the qubit $\mathbb{Q}_1$ that we rotated. As a result,

$$\left.\begin{aligned}\hat{\mathbf{q}}_1(2) &= \left(\hat{\sigma}_x \otimes \hat{1}^4, \quad \left(\cos\theta\hat{\sigma}_y + \sin\theta\hat{\sigma}_z\right) \otimes \hat{1}^2 \otimes \hat{\sigma}_z \otimes \hat{\sigma}_x, \quad \left(\cos\theta\hat{\sigma}_z - \sin\theta\hat{\sigma}_y\right) \otimes \hat{1}^2 \otimes \hat{\sigma}_z \otimes \hat{\sigma}_x\right) \\ \hat{\mathbf{q}}_4(2) &= \left(-\hat{1}^3 \otimes \hat{\sigma}_z \otimes \hat{\sigma}_x, \quad \hat{\sigma}_x \otimes \hat{1}^2 \otimes \hat{\sigma}_y \otimes \hat{\sigma}_x, \quad \hat{\sigma}_x \otimes \hat{1}^2 \otimes \hat{\sigma}_x \otimes \hat{1}\right)\end{aligned}\right\}. \tag{31}$$

Next we use **cnot** gates to perform perfect measurements on $\mathbb{Q}_1$ and $\mathbb{Q}_4$, recording the outcomes in $\mathbb{Q}_2$ and $\mathbb{Q}_3$ respectively:

$$\left.\begin{aligned}\hat{\mathbf{q}}_2(3) &= \left(\hat{1} \otimes \hat{\sigma}_x \otimes \hat{1}^3, \quad \left(\sin\theta\hat{\sigma}_y - \cos\theta\hat{\sigma}_z\right) \otimes \hat{\sigma}_y \otimes \hat{1} \otimes \hat{\sigma}_z \otimes \hat{\sigma}_x, \quad \left(\sin\theta\hat{\sigma}_y - \cos\theta\hat{\sigma}_z\right) \otimes \hat{\sigma}_z \otimes \hat{1} \otimes \hat{\sigma}_z \otimes \hat{\sigma}_x\right) \\ \hat{\mathbf{q}}_3(3) &= \left(\hat{1}^2 \otimes \hat{\sigma}_x \otimes \hat{1}^2, \quad -\hat{\sigma}_x \otimes \hat{1} \otimes \hat{\sigma}_y \otimes \hat{\sigma}_x \otimes \hat{1}, \quad -\hat{\sigma}_x \otimes \hat{1} \otimes \hat{\sigma}_z \otimes \hat{\sigma}_x \otimes \hat{1}\right)\end{aligned}\right\}. \tag{32}$$

(The fact that in this simplified example the information about θ at $t = 2$ is absent from $\mathbb{Q}_4$, and that it is then ($t \geq 3$) carried only in $\mathbb{Q}_2$ and not $\mathbb{Q}_3$, has no fundamental significance: had we been teleporting a general pure state, which would require us to choose two real parameters at A instead of one, $\hat{\mathbf{q}}_1(2)$, $\hat{\mathbf{q}}_4(2)$, $\hat{\mathbf{q}}_2(3)$ and $\hat{\mathbf{q}}_3(3)$ would all generically depend on both those parameters, and both $\mathbb{Q}_2$ and $\mathbb{Q}_3$ would be needed to transport the information about our choice to B.)

Next we subject $\mathbb{Q}_2$, $\mathbb{Q}_3$ and $\mathbb{Q}_5$ to the special transformation **T**:

$$\text{'}\mathbf{T}\text{'}: \quad \begin{Bmatrix}\hat{\mathbf{q}}_k(t+1) \\ \hat{\mathbf{q}}_l(t+1) \\ \hat{\mathbf{q}}_m(t+1)\end{Bmatrix} = \begin{vmatrix}\left(-\hat{q}_{kx}(t)\hat{q}_{mx}(t), & -\hat{q}_{ky}(t)\hat{q}_{mx}(t), & \hat{q}_{kz}(t)\right) \\ \left(\hat{q}_{kz}(t)\hat{q}_{lx}(t)\hat{q}_{mz}(t), & \hat{q}_{kz}(t)\hat{q}_{ly}(t)\hat{q}_{mz}(t), & \hat{q}_{lz}(t)\right) \\ \left(-\hat{q}_{lz}(t)\hat{q}_{mx}(t), & \hat{q}_{kz}(t)\hat{q}_{lz}(t)\hat{q}_{my}(t) & -\hat{q}_{kz}(t)\hat{q}_{mz}(t)\right)\end{vmatrix}, \tag{33}$$

which, as explained by Bennett *et al.* (1993), amounts to performing one of four unitary transformations on $\mathbb{Q}_m$, depending on the binary number stored in $\mathbb{Q}_k$ and $\mathbb{Q}_l$. For present purposes we need only consider the net effect on $\mathbb{Q}_5$, which is to set

$$\hat{\mathbf{q}}_5(4) = \left(\hat{\sigma}_x \otimes \hat{1} \otimes \hat{\sigma}_z \otimes \hat{1} \otimes \hat{\sigma}_z, \quad \left(\cos\theta\hat{\sigma}_y + \sin\theta\hat{\sigma}_z\right) \otimes \hat{\sigma}_z \otimes \hat{\sigma}_z \otimes \hat{\sigma}_z \otimes \hat{\sigma}_z, \quad \left(\cos\theta\hat{\sigma}_z - \sin\theta\hat{\sigma}_y\right) \otimes \hat{\sigma}_z \otimes \hat{1} \otimes \hat{\sigma}_z \otimes \hat{1}\right). \tag{34}$$

Teleportation is now (at $t = 4$) complete. To verify this, note first that $\mathbb{Q}_5$ is now in a pure state – *i.e.* it is no longer entangled with anything. In the Heisenberg picture, the condition that a qubit is pure (given that the overall Heisenberg state is pure) is that there exist a Boolean observable (*i.e.* a projection operator) on that qubit whose measurement is guaranteed to have the outcome 1. This condition is satisfied by $\mathbb{Q}_5$ at $t = 4$, since

$$\left\langle \tfrac{1}{2}\left(\hat{1} - \sin\theta \hat{q}_{5y}(4) - \cos\theta \hat{q}_{5z}(4)\right)\right\rangle = 1. \tag{35}$$

A necessary and sufficient condition for the teleportation to have been successful is that the probability of each possible outcome of each possible measurement on $\mathbb{Q}_5$ at $t = 4$ be the same as the probability of the same outcome of the corresponding measurement on $\mathbb{Q}_1$ at $t = 1$ (just after we rotated $\mathbb{Q}_1$ through the arbitrary angle $\theta$). Since $\mathbb{Q}_5$ is un-entangled, it suffices to consider measurements on it alone, and so, since

$$\left\langle \hat{\mathbf{q}}_1(1)\right\rangle = \left\langle \hat{\mathbf{q}}_5(4)\right\rangle = \left(0, \ -\sin\theta, \ -\cos\theta\right), \tag{36}$$

the condition is met.

Experimentally, one would verify that the information about $\theta$ has reached $\mathbb{Q}_5$ by rotating $\mathbb{Q}_5$ through an angle $-\theta$ about its *x*-axis, after which

$$\hat{q}_{5z}(5) = \cos^2\theta \hat{\sigma}_z \otimes \hat{\sigma}_z \otimes \hat{1} \otimes \hat{\sigma}_z \otimes \hat{1} + \cos\theta \sin\theta \hat{\sigma}_y \otimes \hat{\sigma}_z \otimes \left(\hat{\sigma}_z \otimes \hat{\sigma}_z \otimes \hat{\sigma}_z - \hat{1} \otimes \hat{\sigma}_z \otimes \hat{1}\right) + \sin^2\theta \hat{\sigma}_z \otimes \hat{\sigma}_z \otimes \hat{\sigma}_z \otimes \hat{\sigma}_z \otimes \hat{\sigma}_z, \tag{37}$$

and then measuring whether $\mathbb{Q}_5$ holds the value 0. The probability that it does is predicted to be $\tfrac{1}{2}\left\langle \hat{1} - \hat{q}_{5z}(5)\right\rangle = 1$.

Once again, we see exactly how the information about the angle $\theta$ reached B: not through 'nonlocal influences' allowing it to 'fly across the entanglement' (Jozsa (1998)); not by residing in $\mathfrak{R}$ as a whole rather than in any particular qubit (Braunstein (1996)); not by travelling backwards in time to $t = 1$ with $Q_4$ and then forwards again with $Q_5$ (Penrose (1998)); not instantaneously (a traditional misconception that has sometimes found its way into textbooks – e.g. Williams and Clearwater (1998, §8.10)), nor through action at a distance (Williams and Clearwater (1998, §9.2)); nor of course through the 'collapse of the state vector' (since the state vector is strictly constant) – but simply, prosaically, in the qubits $Q_2$ and $Q_3$ as they travelled from A to B.

## 6. Locally inaccessible information

Qubits $Q_2$ and $Q_3$ do not contain a copy of *all* the information in $Q_1$ and $Q_4$, but only that which has survived decoherence (in the computation basis). Therefore, to local experiments, $Q_2$ and $Q_3$ look like a classical information channel through which the four possible outcomes of the 'Bell measurement' that took place between $t = 1$ and $t = 3$ are transmitted from A to B. However, as we have just seen, this channel also carries a qubit's worth of quantum information, which is locally inaccessible while in transit. This information is transmitted extremely reliably by the decoherent channel, arriving intact provided only that there is no error in communicating the classical message. This illustrates an interesting tradeoff between accessibility and robustness for quantum information: In its simplest manifestations (say, in a single qubit prepared in a pure state) all the quantum information is locally accessible, but it is also maximally vulnerable to decoherence. In contrast, the quantum information that travels from A to B in the teleportation experiment is invulnerable to decoherence but absolutely inaccessible to local experiments.

This tradeoff, which is to be expected given that decoherence processes can be regarded as measurements of the quantum system by the environment (Zurek (1981)), shows us the true role of entanglement in quantum teleportation: entanglement provides a *key* that determines when and how quantum information can be extracted from a decoherent channel. Thus in our teleportation experiment, the inverse Bell operation at $t=0$ sets up algebraic relationships (such as $\langle \hat{q}_{4z}(1)\hat{q}_{5z}(1)\rangle \neq \langle \hat{q}_{4z}(1)\rangle\langle \hat{q}_{5z}(1)\rangle$) between $\hat{\mathbf{q}}_4(t \geq 1)$ and $\hat{\mathbf{q}}_5(t \geq 1)$. These relationships constitute quantum information that is not locally accessible in either $\mathbb{Q}_4$ or $\mathbb{Q}_5$, and is the key that is copied into $\mathbb{Q}_2$ and $\mathbb{Q}_3$ by the measurements at $t=2$, and then allows $\mathbb{Q}_5$ to recover the quantum information about θ that is hidden in $\mathbb{Q}_2$ and $\mathbb{Q}_3$. This may be contrasted with existing interpretations of quantum teleportation, where it is the *classical* information transmitted in qubits $\mathbb{Q}_2$ and $\mathbb{Q}_3$ that would be interpreted as the key, while entanglement is deemed to provide a channel that is neither material nor located in spacetime but through which, nevertheless, the quantum information somehow passes from A to B.

Consider the moment $t=2$ in our EPR experiment (Section 4), when we have just rotated qubit $\mathbb{Q}_2$ through an angle θ, and suppose that φ = 0. Neither of the regions A or B then contains any locally accessible information about θ, but the composite system still does. This ability to 'store information in the correlations between subsystems' is often misrepresented as a nonlocality property of quantum physics, but in fact it is not a uniquely quantum phenomenon at all. For example, imagine that Alice and Bob share a random string of bits $r = (r_1 r_2 \ldots r_n)$ at time $t=0$, and then move to spatially separated regions A and B. Alice composes a text $x = (x_1 x_2 \ldots x_n)$ and encodes it as $y = x \oplus r$, where $\oplus$ is the bitwise **exclusive-or** operation, and then discards the original. As a result the text $x$ is not retrievable from region A alone nor, of course, from region B alone, but only from the combined system. Nevertheless, the information about $x$ does not jump out of region A to an indeterminate location when Alice performs her **exclusive-or** operation but is, in the following sense, located entirely at A throughout: $r$ and $y$ are both random numbers, and given only the mathematical relationship $y = x \oplus r$ between them, which is equivalent to $r = x \oplus y$, either of them could be regarded as the cyphertext version of $x$ while the other was the key needed to extract $x$ from that cyphertext. Nevertheless the history of information flow in the combined system is that $y$, and not $r$, was constructed from $x$, and $r$ but not $y$ was constructed independently of $x$. Hence $y$ is genuinely the cyphertext and $r$ genuinely the key, and consequently the information about $x$ is located at A and not elsewhere.

All phenomena that have been thought to demonstrate nonlocality in quantum physics are actually due to the phenomenon of locally inaccessible information. That is to say, what has been mistaken for nonlocality is the ability of quantum systems to store information in a form which, like a cyphertext, is accessible only after suitable interactions with other systems. It is worth noting that not all such phenomena involve entanglement: the discovery by Bennett *et al.* (1998), which they called 'nonlocality without entanglement', must now be understood as a proof that locally inaccessible information can exist even in non-entangled quantum systems.

Returning now to our EPR experiment with $\phi = 0$, we note that at $t = 2$, all of the network's information about $\theta$ is localised in $\mathbb{Q}_2$. It is locally inaccessible there, but accessible in $\mathbb{Q}_2$ and $\mathbb{Q}_3$ jointly. Thus, again, $\mathbb{Q}_3$ holds the 'key' for accessing the information about $\theta$ in $\mathbb{Q}_2$. Given (22), the Schrödinger state at $t = 2$ is

$$|\psi(2)\rangle = \left(e^{-i\frac{\theta}{2}\hat{\sigma}_x} \otimes \hat{1}\right)|\psi(1)\rangle = \frac{1}{\sqrt{2}}\left[\sin\tfrac{1}{2}\theta\left(|1\rangle|0\rangle - |0\rangle|1\rangle\right) + i\cos\tfrac{1}{2}\theta\left(|0\rangle|0\rangle - |1\rangle|1\rangle\right)\right]. \quad (38)$$

But it is easy to verify that

$$\left(e^{-i\frac{\theta}{2}\hat{\sigma}_x} \otimes \hat{1}\right)|\psi(1)\rangle = \left(\hat{1} \otimes e^{+i\frac{\theta}{2}\hat{\sigma}_x}\right)|\psi(1)\rangle, \quad (39)$$

which means that the Schrödinger state would have been exactly the same if we had placed the information about $\theta$ in $\mathbb{Q}_3$ instead of $\mathbb{Q}_2$, by rotating $\mathbb{Q}_3$ through an angle $-\theta$ – just as, in our classical example, we could have obtained $y$ and $r$ with the same probability distribution function by first choosing $y$ randomly and then constructing $r$ from $x$ and $y$. It follows that in general, to determine where the information about a given parameter is located at a given instant, it is insufficient to know how the Schrödinger state at that instant depends on the parameter. (In contrast, as we have seen, it *is* sufficient to know how the Heisenberg observables at that instant depend on the parameter.)

## 7. Irrelevance of Bell's theorem

Some readers may be hearing a warning Bell in their minds at the idea that the purely local accounts given in Sections 4 and 5 above – or any purely local account – can be compatible with predictions of quantum theory such as (28) and (36). Such readers will be considering *reductio-ad-absurdum* proofs that supposedly rule out all such accounts, along the following lines:

Suppose that at $t = 3$ in our EPR experiment we allow the Boolean observables $\hat{z}_1(3)$ and $\hat{z}_4(3)$ to be measured by observers at A and B respectively, and suppose that the outcomes $a$ and $b$ of these measurements are determined by some local stochastic processes that select each actual outcome from the possibilities $\{0,1\}$. Since the angles $\theta$ and $\phi$ were chosen after the qubits were separated, the effective content of the locality condition is that the stochastic Boolean variables $a$ and $b$ must be independent of $\phi$ and $\theta$ respectively.

For the stochastic processes determining $a$ and $b$ to be consistent with the probabilistic predictions of quantum theory, we must have

$$\left.\begin{aligned}\overline{a(\theta)} &= \langle \hat{z}_1(3) \rangle = \tfrac{1}{2}\left(1 + \langle \hat{\sigma}_z \otimes (\cos\theta \hat{\sigma}_z - \sin\theta \hat{\sigma}_y) \otimes \hat{\sigma}_x \otimes \hat{1} \rangle\right) = \tfrac{1}{2} \\ \overline{b(\phi)} &= \langle \hat{z}_4(3) \rangle = \tfrac{1}{2}\left(1 - \langle \hat{1} \otimes (\cos\phi \hat{1} \otimes \hat{\sigma}_x + \sin\phi \hat{\sigma}_x \otimes \hat{\sigma}_y) \otimes \hat{\sigma}_z \rangle\right) = \tfrac{1}{2}\end{aligned}\right\} \quad (40)$$

for all $\theta$ and $\phi$, where barred quantities such as $\overline{a(\theta)}$ denote mean values. Furthermore, applying (27) and (4), we obtain

$$\overline{a(\theta)b(\phi)} = \langle \hat{z}_1(3)\hat{z}_4(3)\rangle = \tfrac{1}{2}\sin^2\tfrac{1}{2}(\theta-\phi). \tag{41}$$

From (40) and (41) with $\phi = \theta$, we conclude that for any $\theta$,

$$\overline{(1 - a(\theta) - b(\theta))^2} = 0, \tag{42}$$

and therefore

$$b(\theta) = 1 - a(\theta). \tag{43}$$

Hence from (41) again, for all $\theta_0$ and $\theta_1$,

$$\overline{a(\theta_0)a(\theta_1)} = \tfrac{1}{2}\cos^2\tfrac{1}{2}(\theta_0 - \theta_1). \tag{44}$$

Let $\vee$ denote the logical **or** operation on Boolean variables, so that $p \vee q \equiv p + q - pq$, and set $\theta_j = \tfrac{2\pi}{3} j$ in the identity

$$a(\theta_0) \equiv a(\theta_0)\bigl(a(\theta_1) \vee a(\theta_2)\bigr) + a(\theta_0)\bigl(1 - a(\theta_1) \vee a(\theta_2)\bigr). \tag{45}$$

Then note that $p \vee q$ is itself a stochastic Boolean variable and that such variables are non-negative. Hence, using (44) and (40), we obtain

$$\begin{aligned}
\tfrac{1}{2} &= \overline{a(0)\bigl(a(\tfrac{2\pi}{3}) \vee a(\tfrac{4\pi}{3})\bigr)} + \overline{a(0)\bigl(1 - a(\tfrac{2\pi}{3}) \vee a(\tfrac{4\pi}{3})\bigr)} \\
&\leq \overline{a(0)\bigl(a(\tfrac{2\pi}{3}) + a(\tfrac{4\pi}{3}) - a(\tfrac{2\pi}{3})a(\tfrac{4\pi}{3})\bigr)} + \overline{\bigl(1 - a(\tfrac{2\pi}{3}) - a(\tfrac{4\pi}{3}) + a(\tfrac{2\pi}{3})a(\tfrac{4\pi}{3})\bigr)} \\
&\leq \tfrac{3}{8} - \overline{a(0)a(\tfrac{2\pi}{3})a(\tfrac{4\pi}{3})} \\
&\leq \tfrac{3}{8}
\end{aligned} \tag{46}$$

which is a contradiction. This result is a version (similar to that of Mermin (1985)) of Bell's theorem.

Bell's theorem has often been misinterpreted as implying that the empirical predictions of quantum theory cannot be obtained from *any* local theory (see e.g. d'Espagnat (1971, §11.6)), and hence that quantum theory (and therefore presumably reality as well) has a nonlocal character. In the light of our explicit demonstration that the locality premise is true after all, we must instead infer that another of our premises was at fault. In fact the false premise occurs in the first sentence of the argument, where we assumed that we could assign stochastic variables such as $a(\theta)$ to the 'actual outcomes' of measurements. Comparing this with the general exposition of the quantum theory of computation in Section 2, we notice that no such quantities appear there. It is hardly surprising that assigning a single-valued (albeit stochastic) variable to a physical quantity whose true descriptor is a matrix, soon leads to inconsistency.

Note that despite there being, in general, no *single* 'actual outcome' of a measurement, there is of course a well-defined *set* of actual outcomes (viz. some or all of the eigenvalues of the observable being measured), and a probability for each member of that set. These probabilities are not, however, associated with any stochastic variables – again, no such variables occur in the theory presented in Section 2 – but enter quantum theory through an entirely different, deterministic mechanism (see Deutsch (1999)).

## 8. 'Nonlocality' of the Schrödinger picture

Given that quantum theory is entirely local when expressed in the Heisenberg picture, but appears nonlocal in the Schrödinger picture, and given that the two pictures are mathematically equivalent, are we therefore still free to believe that quantum theory (and the physical reality it describes) is nonlocal?

We are not – just as we should not be free to describe a theory as 'complex' if it had both a simple version and a mathematically equivalent complex version. The point is that a 'local' theory is defined as one for which *there exists* a formulation satisfying the locality conditions that we stated at the end of Section 1 (and a local reality is defined as one that is fully described by such a theory). If we were to classify theories as nonlocal whenever it was *possible* to reformulate them in terms of nonlocal quantities (say, $p+q$ and $p-q$, where $p$ and $q$ are local to A and B respectively), then no theory would qualify as local.

Moreover, although the Schrödinger picture disguises the locality of quantum physical processes, all our results could also, with sufficiently careful analysis, be obtained using the Schrödinger picture. Indeed, although we are not aware of any existing correct analysis of quantum information flow, the Schrödinger picture *has* been used by several authors to reach the bare conclusion that quantum processes are local (e.g. Page (1982); Tipler (1998)). When analysing information flow in the Schrödinger picture it is essential to realise that, as we noted in Section 6, it is impossible to characterise quantum information at a given instant using the state vector alone. To investigate where information is located one must also take into account how that state came about. In the Heisenberg picture this is taken care of automatically, precisely because the Heisenberg picture gives a description that is both complete and local.

Thus the Heisenberg picture makes explicit what is implicit, indeed quite well hidden, in the Schrödinger picture. The latter is optimised for predicting the outcomes of processes given how they were prepared, but (notoriously) not for explaining how the outcomes come about – so it is not surprising that on the face of it, it misrepresents information flow. The relationship between the two pictures is somewhat analogous to that between any descriptive piece of information, such as a text or a digitised image, and an algorithmically compressed version of the same information that eliminates redundancy to achieve a more compact representation. If the compression algorithm used is not 'lossy', then, considered as a description of the original data, the two versions are mathematically equivalent. However, the elimination of redundancy results in strong interdependence between the elements of the compressed description so that, for instance, a localised change in the original data can result in changes all over the compressed version, so that a particular character or pixel from the original is not necessarily located at any particular position in the compressed version. Nevertheless, it would be a serious error to conclude that this 'holistic' property of the compressed description expresses any analogous property in the original text or image, or of course in the reality that they refer to.

**Acknowledgement**

We wish to thank Artur Ekert for suggesting several simplifications and other improvements to the arguments and text of this paper.

**References**
Aspect, A., Dalibard, J., Roger, G. (1982) *Phys. Rev. Lett.* **49** 1804
Barenco, A., Deutsch, D., Ekert, A. and Jozsa, R. (1995) *Phys. Rev. Lett.* **74** 20 4083-6
Bell, J. S. (1964) *Physics* **1** 195-200
Bennett, C.H. and Shor, P.W. (1998) *IEEE Transactions on Information Theory* **44** 6 2724-42
Bennett, C.H., Brassard, G., Crépeau, C., Jozsa, R., Peres, A. and Wootters, W.K. (1993) *Phys. Rev. Lett.* **70** 1895-9
Bennett, C.H., DiVincenzo, D.P., Fuchs, C.A., Mor, T., Rains, E., Shor, P.W., Smolin, J.A. and Wootters, W.K. (1998) preprint quant-ph/980453.
Braunstein, S. L. (1996) *Phys. Rev.* **A53** 3 1900-2
Braunstein, S. L., Mann, A. and Revzen, M. (1992) *Phys. Rev. Lett.* **68** 22 3259-61
Deutsch, D. (1989) *Proc. R. Soc. Lond.* **A425** 73
Deutsch, D. (1999) *Proc. R. Soc. Lond.* (to appear)
Deutsch, D., Hayden, P. and Vedral, V. (1999) (in preparation)
d'Espagnat, B. (1971) *Conceptual Foundations of Quantum Mechanics* W.A. Benjamin
Einstein, A. (1949) quoted in in *Albert Einstein: Philosopher, Scientist*, P.A. Schilpp, Ed., Library of Living Philosophers, Evanston, 3$^{rd}$ edition (1970)
Gottesman, D. (1998) preprint quant-ph/9807006.
Jozsa, R. O. (1998) in *The Geometric Universe* (S. A. Huggett, L. J. Mason, K. P. Tod, S. T. Tsou & N. M. J. Woodhouse, eds), pp. 369-379. Oxford University Press.
Mermin, D. (1985) *Physics Today* **38** 4 38-47
Page, D.N. (1982) *Phys. Lett.* **91A** 2 57-60
Penrose, R. (1998) *Phil. Trans. R. Soc. Lond.* **A356** 1743 1927-39
Tipler, F.J. (1998) (submitted to *Phys. Rev. Lett.*)
Williams, C.P. and Clearwater, S.H. (1998) *Explorations in Quantum Computing* Springer-Verlag New York
Zurek, W. H. (1981) *Phys. Rev.* **D24** 1516-25